\documentclass[12pt,a4paper]{article}
\usepackage{a4wide}
\usepackage{amsmath}
\usepackage{amssymb}
\usepackage{epsfig}
\usepackage{subfigure}
\usepackage{exscale}
\usepackage{float}
\usepackage{bbm}
\usepackage[numbers,sort&compress]{natbib}
\usepackage{pst-plot, pstricks,pst-math}
\usepackage{fancybox,amssymb,color}
\usepackage{graphicx}
\usepackage{pstricks, color, graphicx, epsfig, psfrag}
\usepackage{amsfonts,amsmath,amssymb,slashed}
\usepackage{dsfont}
\usepackage{bbm,bm}
\usepackage{fancyhdr, a4wide}
\usepackage[english]{babel}
\usepackage{subfigure}

\makeatletter
\@addtoreset{equation}{section}
\makeatother

\title{Systematic Effective Field Theory Analysis of the D=2+1 Quantum XY Model at Low Temperatures}

\author{Christoph P.\ Hofmann$^a$ \\ \\
\normalsize{$^a$ Facultad de Ciencias, Universidad de Colima} \\
\vspace{0.3cm}
\normalsize {Bernal D\'iaz del Castillo 340, Colima C.P.\ 28045, Mexico} \\}

\begin{document}

\maketitle

\begin{abstract} \normalsize

The low-temperature properties of the (2+1)-dimensional quantum XY model are studied within the framework of effective Lagrangians up to
three-loop order.
At zero temperature, the system is characterized by a spontaneously broken spin rotation symmetry, O(2) $\to$ 1, where the corresponding
Goldstone bosons are the spin waves or magnons. Even though there is no spontaneously broken order at finite $T$, the low-temperature
behavior of the system is still governed by the spin waves.
The partition function is evaluated and various thermodynamic quantities, including the order parameter, are derived in the presence of a
weak external field. In particular, we show that the spin-wave interaction is repulsive at low temperatures, its magnitude depending on
the strength of the external field. We compare our results with those for the (2+1)-dimensional antiferromagnetic Heisenberg model which,
at $T=0$,
exhibits the spontaneously broken spin rotation symmetry O(3) $\to$ O(2).

\end{abstract}


\maketitle

\section{Introduction}
\label{Intro}

One of the most popular methods to address the low-temperature physics of systems exhibiting collective magnetic behavior is spin-wave
theory. Ferromagnets and antiferromagnets have been analyzed extensively within this framework over the years, both in three and in two
spatial dimensions. In particular, to cope with systems defined in two spatial dimensions, modified spin-wave theory \citep{Tak86,Tak87a}
and spin-wave theory at constant order parameter \citep{KSK03} have been invented.

While spin-wave theory worked well for the Heisenberg model, it was rather unclear whether this method is also readily applicable to the
quantum XY model. It was then shown in
Refs.~\citep{UTT79,GJ87}
that this is indeed the case. Subsequent investigations based on spin-wave theory which address the quantum XY model in two spatial
dimensions at zero temperature, can be found in Refs.~\citep{HOW91,WOH91,Zha93,HHB99}.

In the present work, we are interested in the low-temperature regime of the d=2+1 quantum XY model, where the spin waves are the
relevant degrees of freedom. Our approach, however, is based on an entirely different method -- the method of effective Lagrangians which
universally applies to any system whose low-energy properties are dominated by Goldstone bosons.
Indeed, at zero temperature, the spin waves represent the Goldstone bosons which result from the spontaneously broken internal spin
rotation symmetry of the Heisenberg [O(3) $\to$ O(2)] and the XY model [O(2) $\to$ 1], respectively. It is important to note that, at
finite temperature, spontaneous symmetry breaking in two (or less) spatial dimensions cannot occur due to the Mermin-Wagner theorem
\citep{MW66}. Nonetheless, the spin waves still dominate the physics of the system at low temperatures.

Many articles have addressed the behavior of the quantum XY model in two spatial dimensions near or at the Kosterlitz-Thouless phase
transition, which takes place around $T_{KT} \approx 0.343 J$ for the square lattice \citep{HK98}, $J$ being the exchange coupling. On the
other hand, the thermodynamic properties in the regime $T < T_{KT}$, where the spin-waves are the relevant excitations, have only been
explored scarcely \citep{RR84,Din92,Pir96,SH99,TR01,SFBI01}.
In particular, the effect of the spin-wave interaction on the low-temperature properties of the quantum XY model in two spatial dimensions
has not been considered so far, to the best of our knowledge.

In Refs.~\citep{HL90,HN91,HN93} the thermodynamic behavior of systems with a spontaneously broken global symmetry O($N$) $\to$ O($N$-1)
has been analyzed up to two loops in the effective expansion. In the present study we go one step farther in the low-temperature expansion
by including three-loop effects. The calculation for the general case O($N$) $\to$ O($N$-1), and for two spatial dimensions, has been
presented in Ref.~\citep{Hof10}. That article, however, focused on the special case $N$=3, i.e., on the antiferromagnetic Heisenberg model
and did not discuss the physics of the quantum XY model which is the main topic of the present work.

We emphasize that the results for the quantum XY model presented below do not just correspond to adapting the formulas derived earlier for
general $N$ in Ref.~\citep{Hof10} to the case $N$=2. Rather, we have to renormalize and numerically evaluate a particular contribution in
a three-loop graph which was not relevant for the d=2+1 antiferromagnet. In the d=2+1 quantum XY model this term in fact represents one
of the essential contributions coming from the spin-wave interaction.

The main goal of our systematic three-loop calculation is to answer the question of how the spin-wave interaction manifests itself in
various thermodynamic quantities at low temperatures, and how these results are influenced by an external magnetic or staggered field. As
it turns out, the spin-wave interaction is repulsive at low temperatures, its magnitude depending on the strength of the external field.
It will be very instructive to compare these findings with those referring to the d=2+1 Heisenberg antiferromagnet. We emphasize that our
systematic three-loop calculation appears to be beyond the reach of conventional methods such as spin-wave theory -- or would require such
an immense effort that it would not be worthwhile trying to solve the problem with microscopic methods.

We then discuss in which parameter regime the series derived in this work are valid. On the one hand, our effective results are restricted
to temperatures low compared to the underlying microscopic scale, given by the exchange integral $J$. In the case of the antiferromagnetic
Heisenberg model the domain of validity is also restricted due to the nonperturbatively generated mass gap.

Recently, both the quantum XY model and the antiferromagnetic Heisenberg model in two spatial dimensions have been analyzed numerically in
Refs.~\citep{GHJNW09,GHJPSW11}, where high-precision measurements using the very effective loop-cluster algorithm were presented.
Comparing these numerical results with analytic two-loop calculations obtained within effective field theory, permitted to extract some
combinations of low-energy constants at the permille level -- above all, it demonstrated the correctness of the effective field theory
approach for these systems whose low-energy physics is dominated by
spin waves.

In order to promote the effective Lagrangian method, we would like to provide the interested reader with a series of articles where this
method has been used to address condensed matter problems. These systematic studies include antiferromagnets and ferromagnets in three
\citep{Leu94a,Hof99a,Hof99b,RS99a,RS99b,RS00,Hof02,Hof11a},
two
\citep{HL90,HN91,HN93,Hof10,Hof12a,Hof12b}
and one \citep{GHKW10,Hof12c}
spatial dimensions, as well as two-dimensional antiferromagnets which turn into high-temperature superconductors upon doping
\citep{KMW05,BKMPW06,BKPW06,BHKPW07,BHKMPW07,JKHW09,KBWHJW12,VHJW12}.

The rest of the paper is organized as follows. In Sec.~\ref{EffectiveTheory} we provide a brief outline of the effective Lagrangian
description of the quantum XY model. The evaluation of the partition function up to three-loop order in the low-temperature expansion is
presented in Sec.~\ref{FreeEnergyDensity}. The low-temperature series for various thermodynamic quantities, including the order parameter,
are derived in Sec.~\ref{Results} for the (2+1)-d quantum XY model. In particular, the impact of the external field on the spin-wave
interaction is discussed there. In Sec.~\ref{Comparison} we compare these results with those describing the (2+1)-d Heisenberg
antiferromagnet. Finally, our conclusions are presented in Sec.~\ref{Summary}, and some technical details regarding the effective
description of Heisenberg and XY magnets, as well as the renormalization and numerical evaluation of a particular three-loop graph are
discussed in Appendices \ref{appendixA} and \ref{appendixB}. 

\section{Effective Field Theory for the D=2+1 Quantum XY Model}
\label{EffectiveTheory}

On the microscopic level, the quantum XY model in two spatial dimensions is described by the Hamiltonian
\begin{equation}
{\cal H} = - J \sum_{\langle xy \rangle} (S^1_x S^1_y + S^2_x S^2_y) - {\vec H} \cdot \sum_x {\vec S_x} , \quad J > 0.
\end{equation}
We assume that the lattice is bipartite, where $x$ and $y$ represent nearest-neighbor lattice sites with spacing $a$. $J$ is the
ferromagnetic exchange coupling constant. The spin-$\tfrac{1}{2}$ operators $\vec S_x$ follow the standard commutation relations 
\begin{equation}
[S_x^a,S_y^b] = i \delta_{xy} \varepsilon_{abc} S_x^c .
\end{equation}
Note that, contrary to the Heisenberg model, only the generator $S^3$ commutes with the Hamiltonian. The quantity ${\vec H}=(0,H)$
represents a weak external magnetic field which is restricted to the XY-plane and couples to the magnetization order parameter $\vec S$,
\begin{equation}
{\vec S} = \Big( \sum_x S^1_x, \sum_x S^2_x \Big) .
\end{equation}
At infinite volume and zero temperature, the vacuum expectation value of its second component is different from zero,
\begin{equation}
\Sigma = \langle 0 |\sum_x S^2_x | 0 \rangle / V ,
\end{equation}
indicating that the $O(2)$ spin symmetry is spontaneously broken.

On a bipartite lattice, and in the absence of an external field, the antiferromagnetic quantum XY model is related to the ferromagnetic
one by a unitary transformation. In fact, the mapping also holds if we allow for external fields, as we show now. Applying the similarity
transformation \citep{HOW91}
\begin{equation}
S^1_x \to -S^1_x , \qquad S^2_x \to -S^2_x , \qquad S^3_x \to S^3_x ,
\end{equation}
on every site $x$ of the odd sublattice, the ferromagnetic XY Hamiltonian turns into
\begin{equation}
{\cal H} = J \sum_{\langle xy \rangle} (S^1_x S^1_y + S^2_x S^2_y) - \vec H \cdot \sum_x (-1)^x {\vec S_x} , \qquad J > 0 .
\end{equation}
This Hamiltonian describes the antiferromagnetic XY model in the presence of the field $\vec H$, which we now interpret as staggered field
${\vec H_s}$ in the XY-plane, as it couples to the staggered magnetization order parameter $\sum_x (-1)^x {\vec S_x}$. There is thus a
one-to-one correspondence between the XY ferromagnet in a magnetic field and the XY antiferromagnet in a staggered field on a bipartite
lattice. From now on, in order to compare our results with the Heisenberg antiferromagnet, we stick to "antiferromagnetic XY language".

The systematic effective Lagrangian method is designed for systems which display a spontaneously broken global symmetry. In the case of
the quantum XY model, the continuous spin symmetry O(2) is spontaneously broken by the ground state to 1
at $T=0$.
The antiferromagnetic Heisenberg model, on the other hand, follows the pattern O(3) $\to$ O(2). As a consequence of Goldstone's theorem,
the low-energy dynamics of both systems is governed by the spin-waves or magnons, which are characterized by a linear, i.e., relativistic
dispersion relation.

Essential aspects of the effective Lagrangian technique and the perturbative evaluation of the partition function have been outlined on
various occasions, such that here we only provide a minimum of information needed to understand the present calculation. More details can
be found in Section 2 of Ref.~\citep{Hof10} and in appendix A of Ref.~\citep{Hof11a}. Furthermore, pedagogic introductions to the
effective field theory are provided in Refs.~\citep{Brau10,Bur07,Goi04,Sch03,Leu95,Eck95}.

In the quantum XY model, the relevant excitation at low energies is the magnon. It is convenient to define a unit vector field $U^i(x)$,
\begin{equation}
U^i(x) U^i(x) = 1 , \qquad i=1,2 ,
\end{equation}
which contains the Goldstone boson field in its first component $U^1$. Remember that we have chosen the weak external field to point along
the 2-direction, ${\vec H}_s=(0,H_s)$. Accordingly the spin waves represent fluctuations orthogonal to this direction.

The organization of the effective Lagrangian is based on the number of time and space derivatives the various terms display. Terms with
few derivatives dominate the low-energy physics of the system, while terms containing more derivatives are less important. We are thus
dealing with a systematic expansion in powers of energy and momenta. The leading-order effective Lagrangian of the quantum XY model
contains terms with two time (${\partial}_0 {\partial}_0$) and two space (${\partial}_r {\partial}_r$) derivatives,
\begin{eqnarray}
\label{L2space}
{\cal L}^2_{eff} & = & \mbox{$ \frac{1}{2}$} F_1^2 {\partial}_0 U^i \partial_0 U^i
- \mbox{$ \frac{1}{2}$} F_2^2 \partial_r U^i \partial_r U^i + \Sigma_s H_s^i U^i \nonumber \\
& & \qquad r = 1,2 ,
\end{eqnarray}
as well as a term containing the external staggered field $H_s^i$ and the constant $\Sigma_s$ which is the staggered magnetization at zero
temperature and infinite volume. All these contributions are of momentum order $p^2$ and lead to a relativistic dispersion relation,
\begin{equation}
\omega = \sqrt{v^2 {\vec k}^2 + v^4 M^2} , \qquad v = \frac{F_2}{F_1} ,
\end{equation}
where the quantity $v$ is the spin-wave velocity. If one identifies the spin-wave velocity with the velocity of light, the leading-order
effective Lagrangian becomes (pseudo-) Lorentz invariant. Setting $v \equiv 1$, we may use relativistic notation,
\begin{equation}
\label{L2}
{\cal L}^2_{eff}= \mbox{$ \frac{1}{2}$} F^2 \partial_{\mu} U^i \partial^{\mu} U^i + \Sigma_s H_s^i U^i , \qquad F_1 = F_2 = F .
\end{equation}
The quantity $M$ in the dispersion relation is then interpreted as the "mass" of the magnon, i.e., the energy gap in the spectrum, which
is related to the external field by
\begin{equation}
\label{GBMass}
M^2 = \frac{{\Sigma}_s H_s}{F^2} .
\end{equation}
Note that (pseudo-)Lorentz invariance is an accidental symmetry of ${\cal L}^2_{eff}$, which is not shared by the microscopic XY model.
Moreover, the symmetry only emerges at leading order in the derivative expansion -- higher-order contributions in the effective Lagrangian
explicitly break (pseudo-)Lorentz invariance. In fact, they also break O(2) space rotation symmetry due to the anisotropies of the
underlying lattice which start manifesting themselves at order $p^4$. In the present paper, however, we assume (pseudo-)Lorentz invariance
also at next-to-leading order ${\cal L}^4_{eff}$, and thus obtain \citep{HL90}:
\begin{eqnarray}
\label{Leff4}
{\cal L}^4_{eff} & = & e_1 (\partial_{\mu} U^i \partial^{\mu} U^i)^2 + e_2 (\partial_{\mu} U^i \partial^{\nu} U^i)^2 \nonumber \\
& + & k_1 \frac{\Sigma_s}{F^2} (H_s^i U^i) (\partial_{\mu} U^k \partial^{\mu} U^k)
+ k_2 \frac{{\Sigma}_s^2}{F^4} (H_s^i U^i)^2 \nonumber \\
& + & k_3 \frac{{\Sigma}_s^2}{F^4} H_s^i H_s^i .
\end{eqnarray}
It is important to point out that the conclusions of the present paper are not affected by this idealization, as we explore in the
following sections.

At leading order ($p^2$) we have two coupling constants, $F$ and ${\Sigma_s}$, while at next-to-leading order ($p^4$) five constants,
$e_1, e_2, k_1, k_2$ and $k_3$, are required to describe the (anti)ferromagnetic quantum XY model within effective field theory. In
contrast to the derivative structure of the terms in the effective Lagrangian, symmetry does not fix these coupling constants which
parametrize the physics of the underlying microscopic quantum XY model. Rather, they have to be determined experimentally or in a
numerical simulation. Using magnetic terminology, the square of the effective coupling constant $F$ is the spin stiffness or helicity
modulus. While the quantities ${\Sigma}_s$ and $H^i_s$ represent the staggered magnetization and the staggered external field in the case
of the antiferromagnetic XY model, we may also interpret them as magnetization and external magnetic field of the ferromagnetic XY model,
provided that the underlying microscopic lattice is bipartite.

In Appendix \ref{appendixA} we show that on the effective level as well, there is a one-to-one mapping between the ferromagnetic quantum
XY model in a magnetic field and the antiferromagnetic quantum XY model in a staggered field for bipartite lattices. Note that there is no
such mapping between the ferromagnetic and antiferromagnetic Heisenberg model. Indeed, as is well-known, the Heisenberg ferromagnet is
different, the magnons displaying a quadratic dispersion relation. This essential difference between the quantum XY and the Heisenberg
model, as outlined in Appendix \ref{appendixA}, can be traced back to the nature of the spontaneously broken symmetry: while the group
O(2) is Abelian, the group O(3) is non-Abelian.

\section{Free Energy Density up to Three-Loop Order}
\label{FreeEnergyDensity}

The partition function for (2+1)-dimensional (pseudo-)Lorentz-invariant systems with a spontaneously broken global symmetry O($N$) $\to$
O($N$-1) [at $T=0$] has been evaluated within the effective Lagrangian framework up to three-loop order in Ref.~\citep{Hof10}. For the
Heisenberg antiferromagnet ($N$=3), the rather nontrivial part performed in that reference was the renormalization and numerical
evaluation of a specific three-loop graph. In the present case of the quantum XY model ($N$=2) a different three-loop contribution, as
will become clear below, emerges which needs to be renormalized and evaluated numerically. This quite elaborate calculation, not needed
for the antiferromagnet, is presented in detail in Appendix \ref{appendixB}, as it is rather technical.

We briefly review the essential results derived in Ref.~\citep{Hof10}, required for the subsequent discussion. However, in order not to be
repetitive, we do not review the evaluation of Feynman graphs at finite temperature. The interested reader may consult Ref.~\citep{Hof10}
or appendix A of Ref.~\citep{Hof11a} and the various references given therein.

\begin{figure}
\includegraphics[width=15cm]{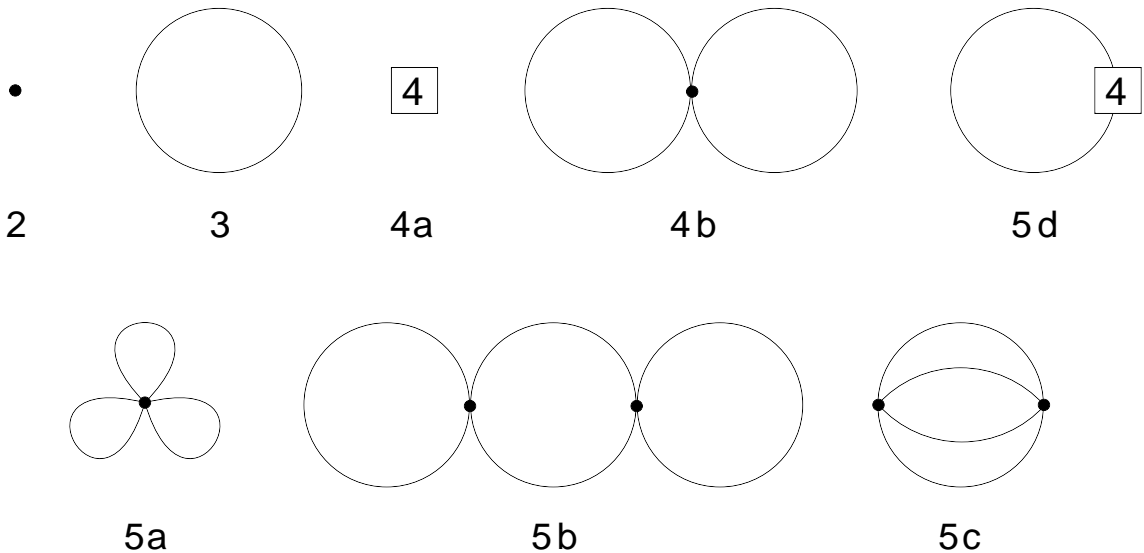}
\caption{Feynman diagrams referring to the low-temperature expansion of the partition function of d=2+1 (pseudo-)Lorentz-invariant systems
with a spontaneously broken symmetry O($N$) $\to$ O($N$-1) up to three-loop order. Vertices associated with the leading term in the
effective Lagrangian, ${\cal L}^2_{eff}$, are denoted by a filled circle, while vertices involving ${\cal L}^4_{eff}$ are referred to by
the number four.}
\label{figure1}
\end{figure}

Up to three-loop order, the eight Feynman diagrams depicted in Fig.~\ref{figure1} are relevant.
The sum of these contributions leads to the following expression for the free energy density for general $N \geq 2$,
\begin{eqnarray}
\label{fedTau}
z & = & z_0 - \frac{N-1}{2} h_0(\sigma) T^3
+ \frac{(N-1)(N-3)}{8 F^2 {\tau}^2} {h_1(\sigma)}^2 T^4 \nonumber \\
& - & \frac{(N-1)(N-3)(5N-11)}{128 \pi F^4 {\tau}^3} {h_1(\sigma)}^2 T^5
+ \frac{(N-1)(N-3)(3N-7)}{48 F^4 {\tau}^2} {h_1(\sigma)}^3 T^5\nonumber \\
& - & \frac{(N-1){(N-3)}^2}{16 F^4 {\tau}^4} {h_1(\sigma)}^2 h_2(\sigma) T^5
+ \frac{1}{F^4} q(\sigma) T^5 + {\cal O}(T^6) ,
\end{eqnarray}
where $z_0$ represents the free energy density at zero temperature. The dimensionless quantities $\sigma$ and $\tau$,
\begin{equation}
\sigma = \frac{M_{\pi}}{2 \pi T} , \qquad \tau = \frac{T}{M_{\pi}} ,
\end{equation}
involve the renormalized mass $M_{\pi}$ of the pseudo-Goldstone bosons,
\begin{eqnarray}
M^2_{\pi} & = & M^2 - \frac{N-3}{8 \pi} \frac{M^3}{F^2} \\
& + & \Big\{ 2 (k_2 - k_1) + \frac{b_1}{F^2} + \frac{b_2}{64 \pi^2 F^2}
\Big\} \frac{M^4}{F^2} + {\cal O}(M^5) , \nonumber 
\end{eqnarray}
which contains higher-order corrections to the leading term
\begin{equation}
M^2 = \frac{{\Sigma}_s H_s}{F^2} .
\end{equation}
The renormalized mass $M_{\pi}$ thus depends on the external field $H_s$,
\begin{eqnarray}
\label{MassRenormalization}
M^2_{\pi}(H_s) & = & \frac{{\Sigma}_s H_s}{F^2} - \frac{N-3}{8 \pi} \frac{{\Sigma}^{3/2}_s H_s^{3/2}}{F^5} \nonumber \\
& + & \Big\{ 2 (k_2 - k_1) + \frac{b_1}{F^2} + \frac{b_2}{64 \pi^2 F^2}
\Big\} \frac{{\Sigma}^2_s H_s^2}{F^6} \nonumber \\
& + & {\cal O}(H_s^{5/2}) .
\end{eqnarray}
The coefficients $b_1$ and $b_2$ are given by
\begin{eqnarray}
b_1 & = & \mbox{$\frac{1}{24}$} (N-3) \gamma_2 - \mbox{$\frac{1}{2}$} (N-2)  \gamma_4 , \nonumber \\
b_2 & = & (N-3)(2N-5) .
\end{eqnarray}
The quantities $\gamma_2$ and $\gamma_4$ are singular functions of the space-time dimension $d$ -- the explicit expressions can be found
in Ref.~\citep{Hof10}. The essential point is that the infinities in $b_1$, which originate from the three-loop graph 5c, are removed by
the combination $k_2-k_1$ of next-to-leading-order coupling constants, stemming from the one-loop graph 5d. As a consequence, the curly
bracket in Eq.~(\ref{MassRenormalization}) is free of singularities.

The dimensionless kinematical functions $h_0, h_1$ and $h_2$, defined by
\begin{equation}
\label{ThermalDimensionless}
g_0 = T^3 h_0(\sigma) , \qquad g_1 = T h_1(\sigma) , \qquad g_2 = \frac{1}{T} h_2(\sigma) ,
\end{equation}
are related to the functions $g_r(M,T)$ which represent Bose functions in $d$ space-time dimensions,
\begin{eqnarray}
\label{BoseFunctions}
g_r(M,T) & = & 2 {\int}_{0}^{\infty} \frac{\mbox{d} \rho}{{(4 \pi \rho)}^{d/2}} {\rho}^{r-1} \exp(- \rho M^2) \nonumber \\
& & \times \sum_{n=1}^{\infty} \exp(-n^2 / 4 \rho T^2) .
\end{eqnarray}
The dimensionless quantity $q$,
\begin{equation}
\label{definitionI}
T^5 q(\sigma) \equiv \mbox{$ \frac{1}{48}$} (N-1)(N-3) M_{\pi}^4 {\bar J}_1 - \mbox{$ \frac{1}{4}$} (N-1) (N-2) {\bar J}_2 ,
\end{equation}
involves the renormalized integrals ${\bar J}_1$ and ${\bar J}_2$,
\begin{eqnarray}
\label{decompositionJ}
{\bar J}_1 & = & J_1 - c_1 - c_2 g_1(M,T) , \nonumber \\
{\bar J}_2 & = & J_2 - c_3 - c_4 g_1(M,T) , 
\end{eqnarray}
which originate from the three-loop graph 5c. The unrenormalized and thus singular integrals $J_1$ and $J_2$ are given by
\begin{eqnarray}
J_1 & = & {\int}_{{\cal T}} {\mbox{d}}^d x \Big\{ G(x) \Big\}^4 ,
\nonumber \\
J_2 & = & {\int}_{{\cal T}} {\mbox{d}}^d x \Big\{ \partial_{\mu} G(x) \partial_{\mu} G(x) \Big\}^2 .
\end{eqnarray}
Here, $G(x)$ is the thermal propagator,
\begin{equation}
\label{ThermalPropagator}
G(x) = \sum_{n = - \infty}^{\infty} \Delta({\vec x}, x_4 + n \beta) ,
\end{equation}
and $\Delta(x)$ represents the zero-temperature Euclidean propagator,
\begin{eqnarray}
\label{regprop}
\Delta (x) & = & (2 \pi)^{-d} \int {\mbox{d}}^d p e^{ipx} (M^2 + p^2)^{-1} \nonumber \\
& = & {\int}_{0}^{\infty} \mbox{d} \rho (4 \pi \rho)^{-d/2} e^{- \rho M^2 - x^2/{4 \rho}} ,
\end{eqnarray}
dimensionally regularized in the space-time dimension $d$. The coefficients $c_1, \dots c_4$ in Eq.~(\ref{decompositionJ}) are subtraction
constants which absorb the infinities in the integrals $J_1$ and $J_2$ -- the explicit representations can be found in appendix A of
Ref.~\citep{Hof10} and in appendix \ref{appendixB} of the present article. Inspecting Eq.~(\ref{definitionI}), it becomes evident that
while the contribution ${\bar J}_2$ is relevant for the Heisenberg model, for the quantum XY model it is the contribution ${\bar J}_1$
that matters -- this is the new expression we have to renormalize and evaluate numerically (see appendix \ref{appendixB}).

For general $N \geq 2$, according to Eq.~(\ref{fedTau}), the leading contribution in the free energy density is of order $T^3$ and
represents the free Bose gas term, corresponding to the one-loop graph 3. The dominant interaction term is of order $T^4$ and originates
from the two-loop graph 4b. At order $T^5$ we have a total of three three-loop graphs that contribute to the interaction, all of them
involving the leading-order Lagrangian ${\cal L}^2_{eff}$ only. Note that we are dealing with a series characterized by integer powers of
the temperature, the leading contribution of order $T^3$ receiving corrections of ascending powers of $T$. This is a consequence of
the fact that (i) each additional loop in a Feynman diagram corresponds to one additional power of $p \propto T$ in two spatial dimensions
and that (ii) the
spin waves
are characterized by a linear dispersion relation.

For the d=2+1 quantum XY model ($N$=2), the free energy density (\ref{fedTau}) reads
\begin{eqnarray}
\label{fedTauN2}
z^{\mbox{\tiny XY}} & = & z^{\mbox{\tiny XY}}_0 - \mbox{$\frac{1}{2}$} h_0(\sigma) T^3 - \frac{1}{8 F^2 {\tau}^2} h^2_1(\sigma) T^4 \\
& - & \frac{1}{128 \pi F^4 {\tau}^3} h^2_1(\sigma) T^5
+ \frac{1}{48 F^4 {\tau}^2} h^3_1(\sigma) T^5 \nonumber \\
& - & \frac{1}{16 F^4 {\tau}^4} h^2_1(\sigma) h_2(\sigma) T^5 + \frac{1}{F^4} q^{\mbox{\tiny XY}}(\sigma) T^5 + {\cal O}(T^6) . \nonumber
\end{eqnarray}
The leading contribution due to the spin-wave interaction is of order $T^4$, subsequent corrections are of order $T^5$.

Interestingly, for the d=2+1 Heisenberg antiferromagnet ($N$=3), many terms in Eq.~(\ref{fedTau}) vanish and the free energy density takes
the simple form
\begin{equation}
\label{freeEnergyN3}
z^{\mbox{\tiny AF}} = z_0^{\mbox{\tiny AF}} - h_0(\sigma) T^3 + \frac{1}{F^4} q^{\mbox{\tiny AF}}(\sigma) T^5 + {\cal O}(T^6) .
\end{equation}
Here, unlike for the spin waves in the quantum XY model, the interaction starts manifesting itself only at order $T^5$.

For the above low-temperature series to be valid, it is important that the quantities $T$ and $M_{\pi}$ (i.e., the external field $H_s$)
are small compared to the intrinsic scale defined by the underlying microscopic theory. This scale is given by the exchange integrals $J$
of the quantum XY and the Heisenberg model. While the ratios $\sigma = M_{\pi}/2 \pi T$ and $\tau = T/ M_{\pi}$ can take any values in
three spatial dimensions, restrictions are imposed in two spatial dimensions. In particular, as we discuss in the next section, the
external field $H_s$ cannot be switched off completely.

We emphasize that anisotropies due to the geometry of the underlying microscopic lattice only show up at higher orders of the derivative
expansion in the effective theory. The different discrete symmetries of e.g. the square and the honeycomb lattice, only manifest
themselves at order $p^4$ in the effective Lagrangian. These space-anisotropies do not affect the main conclusions of the present paper --
and a (pseudo-)Lorentz-invariant framework, even at next-to-leading order of the derivative expansion, is perfectly justified as we
further underline in the next section.

\section{Low-Temperature Series for the D=2+1 Quantum XY Model}
\label{Results}

We now provide the low-temperature expansions for various thermodynamic quantities, including the order parameter. In particular, we
explore how the strength of the spin-wave interaction depends on the magnitude of the external field and on temperature, and answer the
question in which regions of parameter space, defined by $T$, $M_{\pi}$ (i.e., $H_s$) and $F$, our effective expansions are valid.

\begin{figure}
\includegraphics[width=15cm]{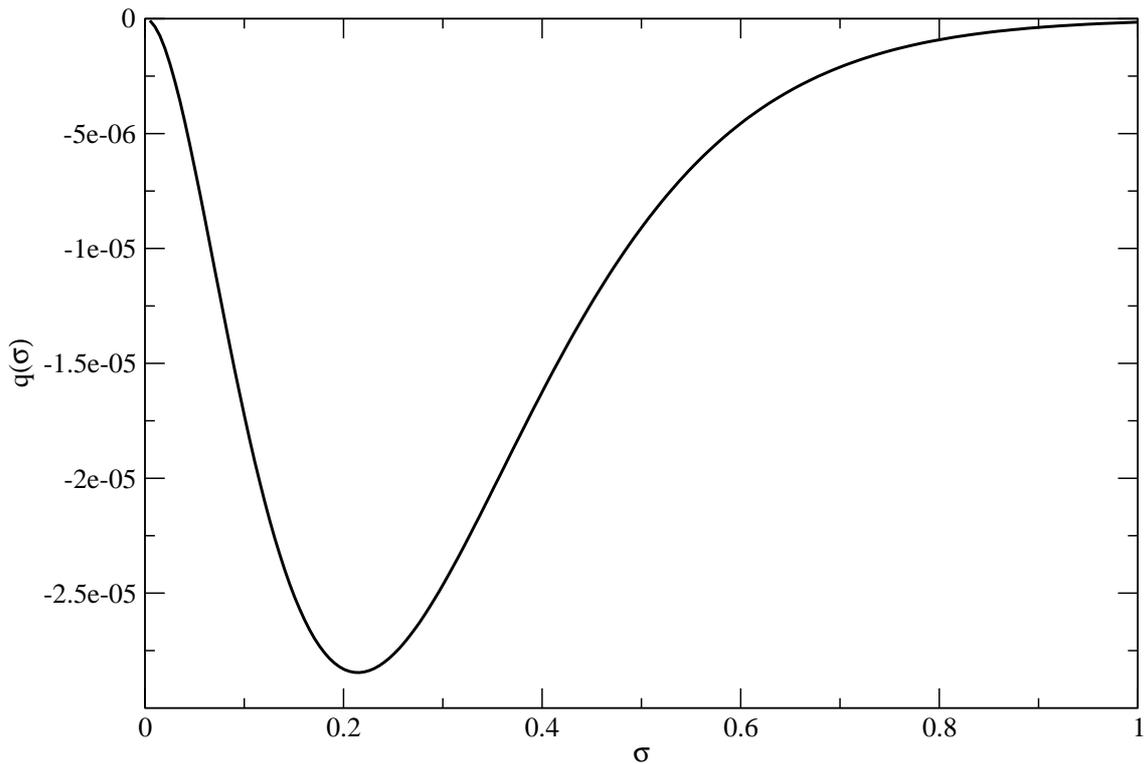}
\caption{The function $q(\sigma)$ for $N$=2, with $\sigma=M_{\pi}/2 \pi T$ as dimensionless parameter.}
\label{figure2}
\end{figure}

Up to three loops, the pressure for the d=2+1 quantum XY model takes the form
\begin{eqnarray}
\label{pressureTauN2}
P^{\mbox{\tiny XY}} & = & z^{\mbox{\tiny XY}}_0 - z^{\mbox{\tiny XY}}
= \mbox{$\frac{1}{2}$} h_0(\sigma) T^3
+ \frac{1}{8 F^2 {\tau}^2} {h_1(\sigma)}^2 T^4 \nonumber \\
& + &\frac{1}{128 \pi F^4 {\tau}^3} {h_1(\sigma)}^2 T^5
- \frac{1}{48 F^4 {\tau}^2} {h_1(\sigma)}^3 T^5 \\
& + & \frac{1}{16 F^4 {\tau}^4} {h_1(\sigma)}^2 h_2(\sigma) T^5
- \frac{1}{F^4} q^{\mbox{\tiny XY}}(\sigma) T^5 + {\cal O}(T^6) . \nonumber
\end{eqnarray}
The function $q(\sigma)$, defined in Eq.~(\ref{definitionI}), is depicted for $N$=2 in Fig.~\ref{figure2}. It develops a minimum around
$\sigma \approx 0.22$ and tends to zero both for small and large values of $\sigma$. Expressions suitable for the numerical evaluation of
$q(\sigma)$ are given in Appendix \ref{appendixB}. Roughly speaking, according to Eq.~(\ref{MassRenormalization}), the parameter
$\sigma = M_{\pi}/2 \pi T$ is proportional to $\sqrt{H_s}/T$, i.e., the ratio between the square root of the external field and the
temperature.

\begin{figure}
\includegraphics[width=15cm]{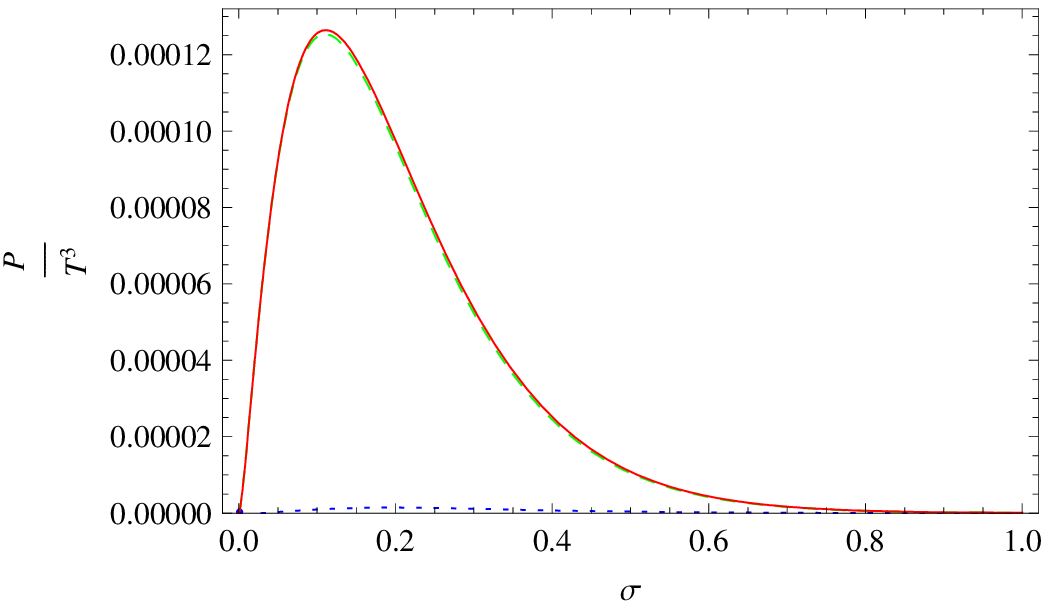}
\caption{(2+1)-dimensional quantum XY model: Interaction corrections to the pressure as a function of the dimensionless ratio
$\sigma=M_{\pi}/2 \pi T$, evaluated at $T/F^2 = \tfrac{1}{2} T_{KT}/J$, where $T_{KT} \approx 0.343 J$ is the Kosterlitz-Thouless
transition temperature. The two-loop contribution (dashed curve) is slightly enhanced by the three-loop contribution (dotted curve).
The sum (continuous curve) develops a maximum around $\sigma = 0.11$.}
\label{figure3}
\end{figure}

\begin{figure}
\includegraphics[width=15cm]{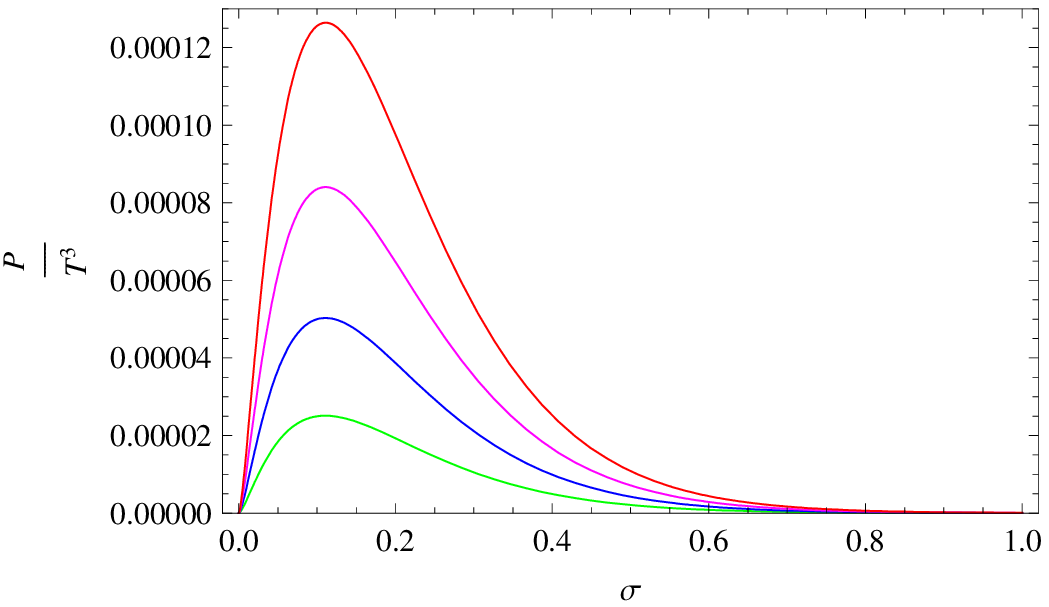}
\caption{(2+1)-dimensional quantum XY model: The sum of the two-loop and three-loop corrections to the pressure as a function of
$\sigma=M_{\pi}/2 \pi T$, evaluated at $T/F^2 = \{ \tfrac{1}{10}, \tfrac{1}{5}, \tfrac{1}{3}, \tfrac{1}{2} \} T_{KT}/J$, from bottom to
top in the figure.}
\label{figure4}
\end{figure}

The leading interaction contribution of order $T^4$ in the pressure is positive, signaling that the interaction between spin waves is
repulsive. The sum of the three-loop corrections of order $T^5$ slightly enhances the repulsive interaction, as can be appreciated in
Fig.~\ref{figure3}. The interaction is strongest at the value $\sigma \approx 0.11$. According to Fig.~\ref{figure4}, this maximum is
basically independent of the ratio $T/F^2$.

Now on the square lattice we have $F^2 = 0.26974(5) J$ \citep{GHJPSW11}, which is close to the Kosterlitz-Thouless transition temperature
$T_{KT} \approx 0.343 J$. Both quantities, $F^2$ and $T_{KT}$, are of the order of the underlying microscopic scale $J$. For our effective
expansions to be valid, the ratio $T/F^2$ must thus be small. The value $\sigma \approx 0.11$, using the first two terms on the RHS of
Eq.~(\ref{MassRenormalization}), corresponds to $H_s/ T^2 \approx 0.38 /J$ for the square lattice. For this ratio of external field versus
temperature squared, the magnitude of the spin-wave interaction in the pressure is largest. Again, these results apply to the quantum XY
model on the square lattice, where the other low-energy constants take the values $\Sigma_s = 0.43561(1)/a^2$ and $v = 1.1347(2) \; Ja$
\citep{GHJPSW11}.

An important comment on (pseudo-)Lorentz invariance and lattice anisotropies is in order here. As we have pointed out, the leading-order
effective Lagrangian ${\cal L}^2_{eff}$ Eq.~(\ref{L2space}) displays an accidental O(2) space rotation symmetry and can thus be written in
a (pseudo-)Lorentz-invariant form. We then have assumed that this symmetry persists at order ${\cal L}^4_{eff}$. This idealization is
indeed well-justified, as the only place where these Lorentz-symmetry breaking effects manifest themselves in the temperature-dependent
quantities, is through the one-loop diagram 5d of Fig.~\ref{figure1}. This specific diagram only affects the renormalized mass and does so
only at next-to-next-to-leading order in the form of the subleading low-energy constants $k_1$ and $k_2$ in
Eq.~(\ref{MassRenormalization}). From this point of view, the differences between e.g. the square and the honeycomb lattice are
negligible, as they correspond to a next-to-next-to-leading order effect which does not affect at all our conclusions.

A crucial point, on the other hand, is to realize that the geometry of the lattice does manifest itself in a rather trivial way. Although
the leading-order effective Lagrangian ${\cal L}^2_{eff}$ is (pseudo-)Lorentz invariant -- again, this is a rigorous statement -- the
actual values of the low-energy constants $\Sigma_s$ (staggered magnetization at zero temperature and zero external field), $v$ (spin-wave
velocity) and $F^2$ (spin stiffness or helicity modulus) depend on the lattice geometry. While the structure of the low-temperature
series, i.e. the specific powers of the temperature involved and the dependence on the external field, is universal for bipartite
lattices, the differences between various lattice geometries show up in the coefficients of these series, as they depend on the low-energy
constants.

To further explore the effect of the spin-wave interaction, we also consider the order parameter. The low-temperature expansion of the
staggered magnetization,
\begin{equation}
\Sigma_s(T,H_s) = - \frac{\partial z}{\partial H_s} ,
\end{equation}
amounts to
\begin{eqnarray}
\label{OPTauN2}
\Sigma_s^{\mbox{\tiny XY}}(T,H_s) & = & \Sigma_s^{\mbox{\tiny XY}}(0,H_s) - \frac{\Sigma_s b}{2 F^2} h_1(\sigma) T
+ \frac{\Sigma_s b}{8 F^4} \Big\{ {h_1(\sigma)}^2 - \frac{2}{\tau^2} h_1(\sigma) h_2(\sigma) \Big\} T^2 \nonumber \\
& + & \frac{\Sigma_s b}{128 \pi F^6} \Big\{ {\frac{3}{2 \tau} h_1(\sigma)^2} - \frac{2}{\tau^3} h_1(\sigma) h_2(\sigma) \Big\} T^3
- \frac{\Sigma_s b}{48 F^6} \Big\{ {h_1(\sigma)}^3 -\frac{3}{\tau^2} {h_1(\sigma)}^2 h_2(\sigma) \Big\} T^3 \nonumber \\
& + & \frac{\Sigma_s b}{16 F^6} \Big\{ \frac{2}{\tau^2} {h_1(\sigma)}^2 h_2(\sigma) - \frac{2}{\tau^4} h_1(\sigma) {h_2(\sigma)^2}
- \frac{1}{\tau^4} {h_1(\sigma)}^2 h_3(\sigma) \Big\} T^3 \nonumber \\
& - &\frac{\Sigma_s b}{8 \pi^2 F^6 \sigma} \frac{\partial q^{\mbox{\tiny XY}}(\sigma)}{\partial \sigma } T^3 + {\cal O}(T^4) .
\end{eqnarray}
The quantity $b$, for general $N$, is defined by
\begin{eqnarray}
\label{defb}
b(H_s) & = & \frac{\partial M^2_{\pi}}{\partial M^2} = 1 - \frac{3 (N-3) \sqrt{\Sigma_s}}{16 \pi F^3} \sqrt{H_s}
+ \frac{2 \tilde k_0 \Sigma_s}{F^4} H_s \nonumber \\
& + & {\cal O}(H_s^{3/2}) , \nonumber \\
& & \tilde k_0 = 2 (k_2 - k_1) + \frac{b_1}{F^2} + \frac{b_2}{64 \pi^2 F^2} .
\end{eqnarray}
Inverting Eq.~(\ref{MassRenormalization}), i.e., expressing the staggered field $H_s$ as a function of the renormalized mass $M_{\pi}$, the
factor $b$ can be rewritten as $b= b(M_{\pi})$. This $M_{\pi}$-dependent representation of $b$ will be used in the subsequent plots.
Finally, the quantity $\Sigma_s^{\mbox{\tiny XY}}(0,H_s)$ in Eq.~(\ref{OPTauN2}), is the zero-temperature order parameter in the presence of
the external field. For general $N$ we have
\begin{eqnarray}
\label{SigmaN}
\Sigma_s(0,H_s) & = & \Sigma_s + \frac{(N-1) \Sigma_s^{3/2}}{8 \pi F^3} \sqrt{H_s} + {\cal O}(H_s) , \nonumber \\
\Sigma_s & = & \Sigma^{\mbox{\tiny XY}}_s(0,0) .
\end{eqnarray}
Again, inverting Eq.~(\ref{MassRenormalization}), $\Sigma_s(0,H_s)$ can be written as a function of $M_{\pi}$. Of course, in the following
plots related to the quantum XY model, we have set $N$=2.

\begin{figure}
\includegraphics[width=15cm]{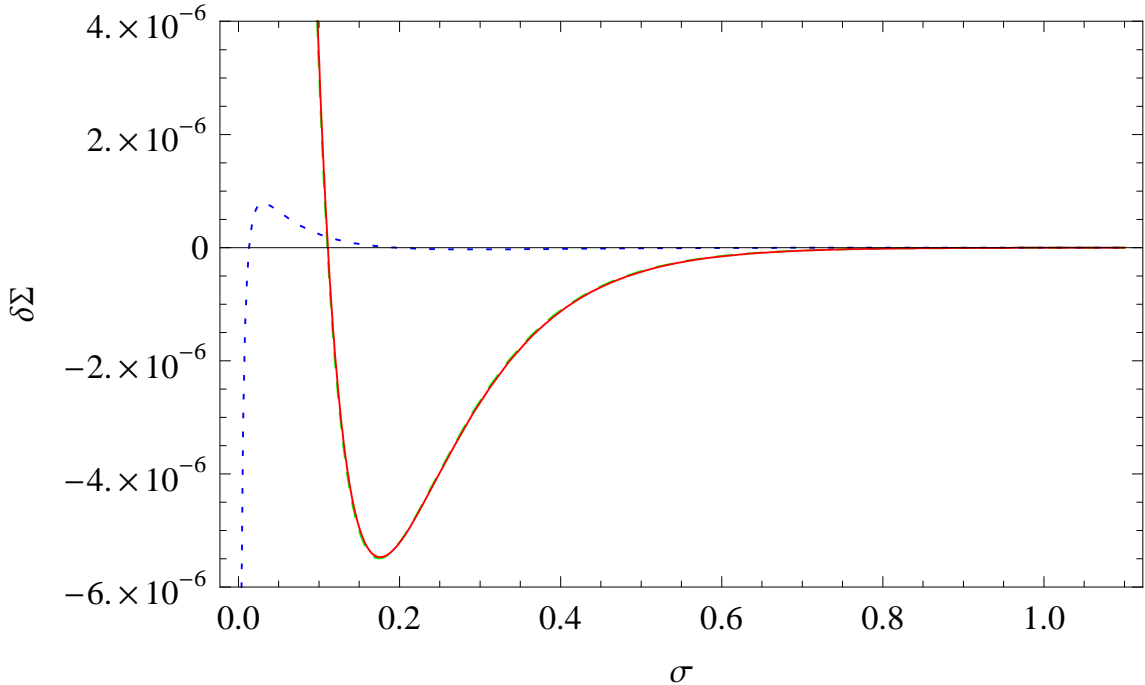}
\caption{(2+1)-dimensional quantum XY model: Temperature-dependent interaction corrections to the staggered magnetization,
$\delta \Sigma_s$, as a function of the dimensionless ratio $\sigma=M_{\pi}/2 \pi T$, evaluated at $T/F^2 = \tfrac{1}{2} T_{KT}/J$. The
two-loop contribution (dashed curve) and three-loop contribution (dotted curve) add up to the total correction (continuous curve) which is
negative in the whole parameter region, except for small values of $\sigma$.}
\label{figure5}
\end{figure}

\begin{figure}
\includegraphics[width=15cm]{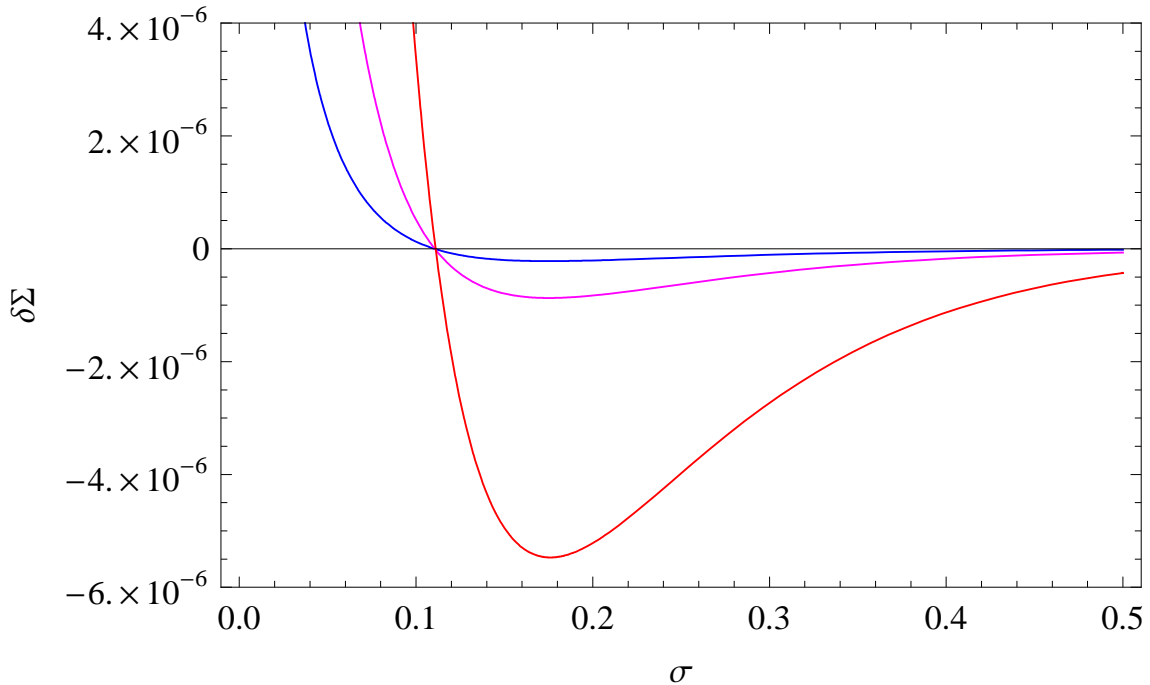}
\caption{(2+1)-dimensional quantum XY model: The sum of the two-loop and three-loop corrections to the staggered magnetization as a
function of the dimensionless ratio $\sigma=M_{\pi}/2 \pi T$, evaluated at $T/F^2 = \{ \tfrac{1}{10},\tfrac{1}{5}, \tfrac{1}{2} \}
T_{KT}/J$, from top to bottom in the figure. The quantity $T_{KT} \approx 0.343 J$ is the Kosterlitz-Thouless phase transition
temperature.}
\label{figure6}
\end{figure}

Our focus is the impact of the spin-wave interaction in the temperature-dependent part of the staggered magnetization, given by the
difference ${\Sigma}_s^{\mbox{\tiny XY}}(T,H_s) - \Sigma_s^{\mbox{\tiny XY}}(0,H_s)$. A plot of the dimensionless quantity
\begin{equation}
\delta \Sigma_s \equiv \frac{\Sigma_s^{\mbox{\tiny XY}}(T,H_s) - \Sigma_s^{\mbox{\tiny XY}}(0,H_s)}{\Sigma_s} ,
\end{equation}
both for the two-loop ($\propto T^2$) and three-loop ($\propto T^3$) contribution in the order parameter Eq.~(\ref{OPTauN2}), is provided
in Fig.~\ref{figure5}. The sum of these contributions is negative in most of the parameter space, except for small values of $\sigma$
where the quantity $\delta \Sigma_s$ is positive. Note that the point where the temperature-dependent spin-wave interaction correction
vanishes, is basically independent of the ratio $T/F^2$, as illustrated in Fig.~\ref{figure6}.

In the low-temperature series derived in the present study, the values of the ratios $T/F^2 \propto T/J$ and $H_s/F^2 \propto H_s/J$ must
be small -- otherwise, we would leave the low-energy domain where the effective expansion applies. However, so far we have not considered
the restrictions imposed by the Mermin-Wagner theorem \citep{MW66}. At finite temperature, in the absence of external fields, there is no
spontaneous symmetry breaking in two spatial dimensions. Taking the limit $M_{\pi} \to 0$ or, equivalently, switching off the external
field in the above low-temperature expansions, is thus problematic.

If we formally take the limit $\sigma \to 0$ in the pressure, internal energy density, entropy density and heat capacity, all interaction
contributions vanish and we are left with the free magnon part given by
\begin{eqnarray}
\label{pressureXY}
P^{\mbox{\tiny XY}} & = & \frac{\zeta(3)}{2 \pi} T^3 + {\cal O}(T^6) , \nonumber \\ 
u^{\mbox{\tiny XY}} & = & \frac{\zeta(3)}{\pi} T^3 + {\cal O}(T^6) , \nonumber \\
s^{\mbox{\tiny XY}} & = & \frac{3 \zeta(3)}{2 \pi} T^2 + {\cal O}(T^5) , \nonumber \\
c^{\mbox{\tiny XY}}_V & = & \frac{3 \zeta(3)}{\pi} T^2 + {\cal O}(T^5) .
\end{eqnarray}
Note that the limit $\sigma = M_{\pi}/2 \pi T \to 0$ is implemented by switching the external field (i.e., $M_{\pi}$) off, while keeping
the temperature finite. Formally, this limit poses no problems in the above quantities. However, analyzing the behavior of the order
parameter, indicates that there is a subtlety, as we now underline.

The leading terms in the low-temperature expansion of the pressure Eq.~(\ref{pressureTauN2}) and the order parameter Eq.~(\ref{OPTauN2})
are proportional to the kinematical function $h_0(\sigma)$ and $h_1(\sigma)$, respectively. Their Taylor expansion in the parameter
$\sigma$, i.e., for a weak external staggered field, takes the form
\begin{eqnarray}
\label{h0h1}
h_0(\sigma) & = & \frac{\zeta(3)}{\pi} - \frac{1}{4 \pi} \frac{\Sigma_s H_s}{F^2 T^2} + \frac{1}{4 \pi} \frac{\Sigma_s H_s}{F^2 T^2} 
\ln \frac{\Sigma_s H_s}{F^2 T^2} + \dots , \nonumber \\
h_1(\sigma) & = & - \frac{1}{4 \pi} \ln \frac{\Sigma_s H_s}{F^2 T^2} + \dots .
\end{eqnarray}
Switching off the external field $H_s$ in the order parameter is thus problematic due to the term $\ln H_s$ in $h_1(\sigma)$ which
becomes divergent. This is precisely where the Mermin-Wagner theorem enters through the backdoor -- the issue is quite subtle. We start by
estimating the temperature where our effective series break down. Consider the first two terms in low-temperature expansion of the order
parameter Eq.~(\ref{OPTauN2}), i.e. the zero-temperature staggered magnetization in presence of the external field $H_s$ and the leading
(one-loop) temperature-dependent contribution. When the staggered magnetization becomes zero, or even takes negative values, the effective
expansion can no longer be trusted. This is because the temperature where the effective analysis operates must be smaller than the
temperature where the spin-wave picture breaks down. For a given value of the staggered field, we thus get an estimate for this "critical"
temperature $T_c$, by solving the equation
\begin{equation}
\label{checkFieldXY}
\Sigma_s^{\mbox{\tiny XY}}(T,H_s) = \Sigma_s + \frac{\Sigma_s^{3/2}}{8 \pi F^3} \sqrt{H_s} - \frac{\Sigma_s b}{2 F^2} h_1(\sigma) T \equiv 0 ,
\end{equation}
where
\begin{equation}
h_1(\sigma) = - \frac{\ln ( 1-e^{-M_{\pi}/T} )}{2 \pi} \approx - \frac{\ln ( 1-e^{-\sqrt{\Sigma_s H_s}/FT} )}{2 \pi} .
\end{equation}
On the square lattice, the ratios $H_s/F^2 = \{ 10^{-1}, 10^{-10}, 10^{-100}, 10^{-1000}, 10^{-10000} \}$ correspond to
$T_c/J = \{ 1.715, 0.3184, 0.03032, 0.002961, 0.0002949 \}$. Weakening the staggered field, the "critical" temperature gradually becomes
smaller, and logarithmically tends to zero in the limit $H_s \to 0$. Hence the temperature range, in which our effective analysis
operates, shrinks to zero. The low-temperature representation of the order parameter, Eq.~(\ref{OPTauN2}), then no longer makes sense --
the staggered field must always be different from zero.

Note that the problematic term $\ln H_s$ in the order parameter, interestingly, in the thermodynamic quantities $P, u, s$ and $c_V$
manifests itself as $H_s \ln H_s$. Although it is also conceptually inconsistent to take the limit $H_s \to 0$ in these quantities,
there is no divergent behavior. In fact, for a given set of $H_s/F^2$ and $T_c$ provided above, the ratio $\Sigma_s H_s/F^2 T_c^2$ is tiny,
such that the first term in the expansion of $h_0$, Eq.~(\ref{h0h1}), dominates over the remainder. As a consequence, neglecting the
nonleading terms altogether does not really numerically affect the low-temperature series for $P, u, s$ and $c_V$. Therefore the results
displayed in Eq.~(\ref{pressureXY}) are correct.

\section{D=2+1 Heisenberg Antiferromagnet at Low Temperatures}
\label{Comparison}

\begin{figure}
\includegraphics[width=11.5cm,angle=-90]{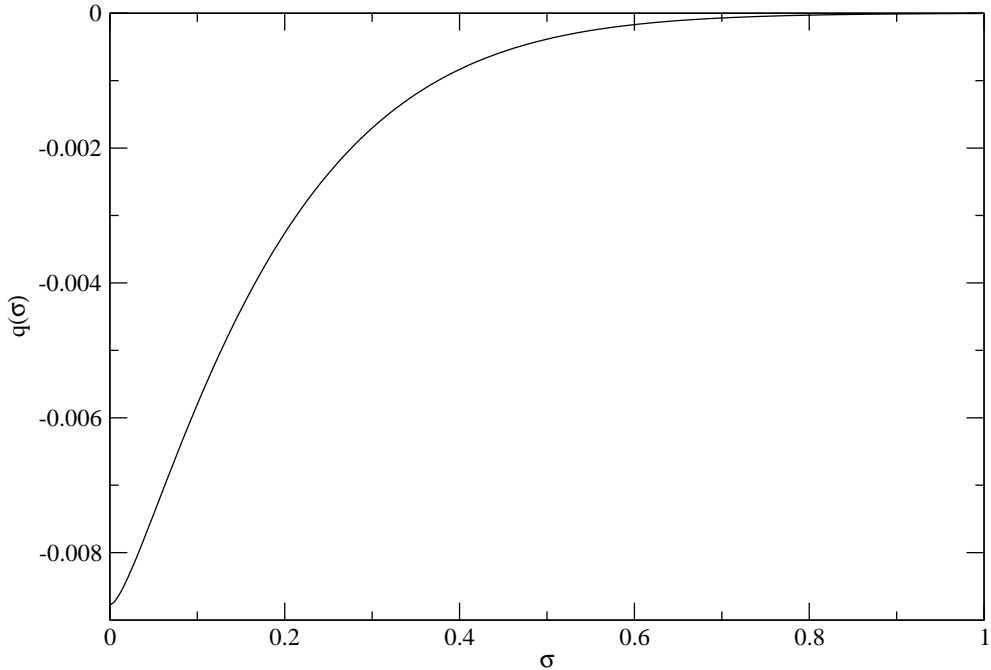}
\caption{The function $q(\sigma)$ for $N$=3, with $\sigma=M_{\pi}/2 \pi T$ as dimensionless parameter.}
\label{figure7}
\end{figure}

In this section we compare the results for the quantum XY model with the analogous results for the d=2+1 Heisenberg antiferromagnet within
our three-loop analysis. Previous effective field theory studies of the d=2+1 Heisenberg antiferromagnet include
Refs.~\citep{CHN88,CHN89,NZ89,Fis89,HN91,HN93}.

\begin{figure}
\includegraphics[width=15cm]{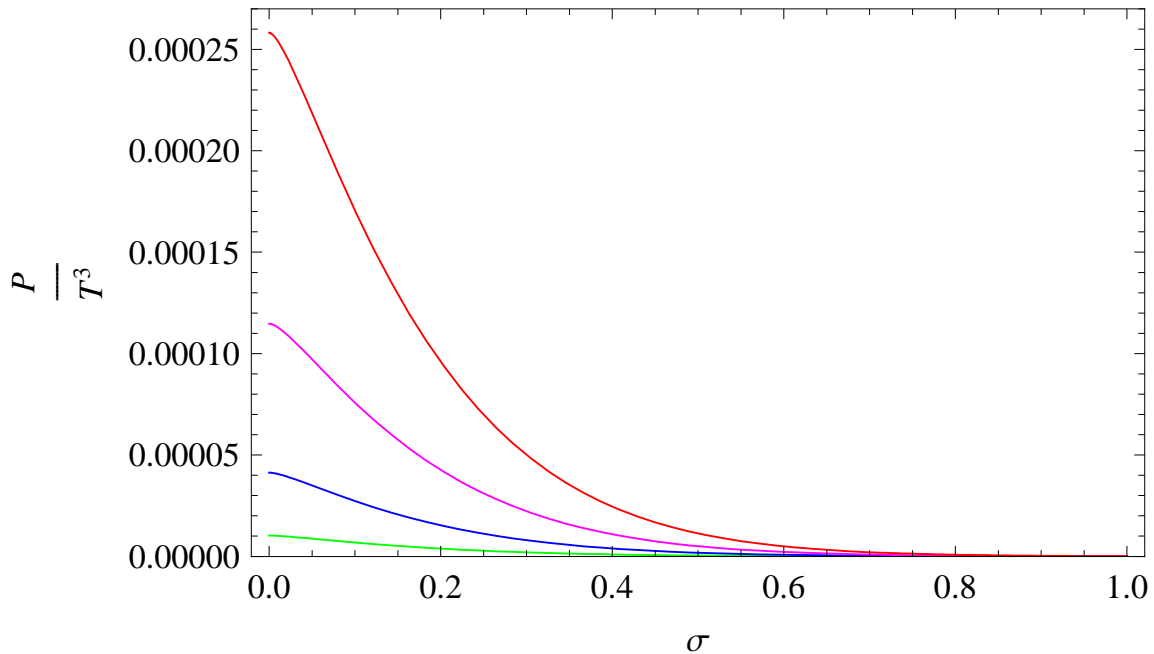}
\caption{(2+1)-dimensional Heisenberg antiferromagnet: Three-loop interaction correction to the pressure as a function of the
dimensionless ratio $\sigma=M_{\pi}/2 \pi T$, evaluated at the same temperatures
$T/F^2 = \{ \tfrac{1}{10}, \tfrac{1}{5}, \tfrac{1}{3}, \tfrac{1}{2} \} T_{KT}/J$ as for the XY model in Fig.~\ref{figure4}, from bottom
to top in the figure.}
\label{figure8}
\end{figure}

Up to three-loop order, the pressure reads
\begin{equation}
\label{Pz}
P^{\mbox{\tiny AF}} = z^{\mbox{\tiny AF}}_0 -z^{\mbox{\tiny AF}} = h_0(\sigma) T^3 - \frac{1}{F^4} q^{\mbox{\tiny AF}}(\sigma) T^5 + {\cal O}(T^6) ,
\end{equation}
with $z^{\mbox{\tiny AF}}$ given in Eq.~(\ref{freeEnergyN3}). A plot for the function $q(\sigma)$, defined in Eq.~(\ref{definitionI}), is
provided for $N$=3 in Fig.~\ref{figure7}. In the whole parameter regime $\sigma$, this function is negative. Accordingly, the spin-wave
interaction in the pressure, much like for the d=2+1 quantum XY model, is always repulsive, as depicted in Fig.~\ref{figure8}.

Interestingly, the spin-wave interaction gets stronger if the staggered field is weakened at fixed temperature -- in contrast to the
quantum XY model, where the strength of the interaction develops a maximum at $\sigma \approx 0.11$. Note that there is no two-loop
contribution in the case of the Heisenberg antiferromagnet. Still, although we are dealing with a three-loop effect, the strength $P/T^3$
of the spin-wave interaction is larger than in the quantum XY model for values of $T/F^2 \geq 0.24$ (compare Fig.~\ref{figure8} with
Fig.~\ref{figure4}). This is because (i) the coefficient of ${\bar J}_1$ in Eq.~(\ref{definitionI}) is a factor of 24 smaller than the one
of ${\bar J}_2$ and (ii) the quantity ${\bar J}_1$ is furthermore suppressed by four powers of $\sigma$, which is small in the relevant
domain considered in the above plots.

\begin{figure}
\includegraphics[width=15cm]{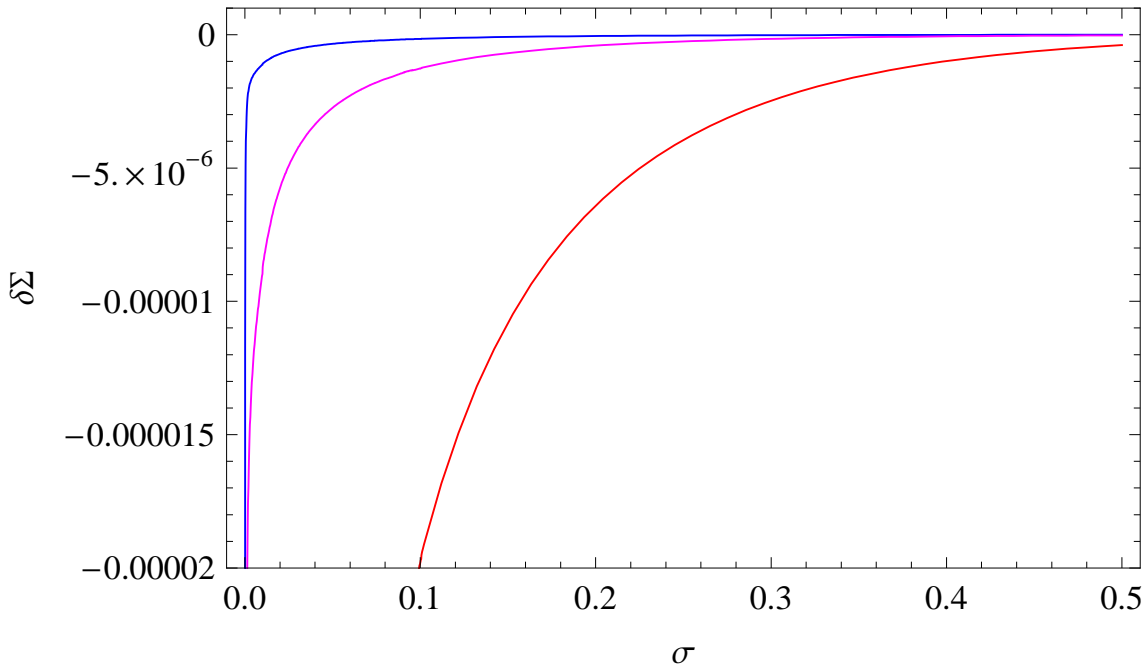}
\caption{(2+1)-dimensional Heisenberg antiferromagnet: Temperature-dependent interaction correction to the staggered magnetization,
$\delta \Sigma_s$, as a function of $\sigma = M_{\pi}/2 \pi T$, evaluated at the same temperatures
$T/F^2 = \{ \tfrac{1}{10}, \tfrac{1}{5}, \tfrac{1}{2} \} T_{KT}/J$ as for the XY model in Fig.~\ref{figure6}, from top to bottom in the
figure. While the two-loop contribution is zero, the three-loop contribution is negative in the whole parameter region.}
\label{figure9}
\end{figure}

In the temperature-dependent part of the staggered magnetization,
\begin{equation}
\delta \Sigma_s \equiv \frac{{\Sigma}_s^{\mbox{\tiny AF}}(T,H_s) - \Sigma_s^{\mbox{\tiny AF}}(0,H_s)}{\Sigma_s} , \qquad
\Sigma_s = \Sigma_s^{\mbox{\tiny AF}}(0,0) ,
\end{equation}
as can be seen in Fig.~\ref{figure9}, the overall sign of the interaction correction is negative in the whole parameter range. Note that
the spin-wave interaction only sets in at three-loop order. In the above plot we have used
\begin{eqnarray}
\label{OPTauN3}
\Sigma_s^{\mbox{\tiny AF}}(T,H_s) & = & \Sigma_s^{\mbox{\tiny AF}}(0,H_s) - \frac{\Sigma_s b}{F^2} h_1(\sigma) T \nonumber \\
& - & \frac{\Sigma_s b}{8 \pi^2 F^6 \sigma} \frac{\partial q^{\mbox{\tiny AF}}(\sigma)}{\partial \sigma } T^3 + {\cal O}(T^4) ,
\end{eqnarray}
along with the definition of $b$ and $\Sigma_s^{\mbox{\tiny AF}}(0,H_s)$ given in Eqs.~(\ref{defb}) and (\ref{SigmaN}) for general $N$.

Formally, in the limit $\sigma \to 0$ (i.e., zero staggered field), the pressure reduces to
\begin{equation}
P^{\mbox{\tiny AF}} = \frac{\zeta(3)}{\pi} T^3 \Big[ 1 - \frac{\pi q_1}{\zeta(3)} \frac{T^2}{F^4} + {\cal O}(T^3) \Big] ,
\end{equation}
where the coefficient $q_1 = q(\sigma = 0)$ is given by $q_1 = -0.008779$ (see Ref.~\citep{Hof10}). In contrast to the quantum XY
model, magnons related to the antiferromagnetic Heisenberg model keep interacting if the staggered external field is weakened. The
corresponding expressions for $u, s$ and $c_V$ can be found in Eq.~(5.7) of Ref.~\citep{Hof10}.

As for the XY model, the staggered magnetization of the Heisenberg antiferromagnet is logarithmically divergent in the limit $H_s \to 0$,
due to the kinematical function $h_1(\sigma)$. Again, this divergence is purely mathematical, as it is forbidden to switch off completely
the staggered field in our effective analysis. In the limit $H_s \to 0$, the "critical" temperature $T_c$, which marks the breakdown of
the spin-wave picture, tends to zero.

But there is an important difference in the low-temperature dynamics of the quantum XY and the AF Heisenberg model in two spatial
dimensions, we want to point out. Unlike the (2+1)-dimensional quantum XY model, the (2+1)-dimensional Heisenberg antiferromagnet develops
a nonperturbatively generated mass gap at finite temperatures \citep{HN90}. As outlined in detail in Ref.~\citep{Hof10}, this subtle
effect also implies that the external field $H_s$ cannot be switched off completely in our effective low-temperature expansions.

Yet another difference between the two models is the following. Although neither one of them exhibits spontaneously broken order at finite
temperature and zero external field, the D=2+1 quantum XY model nonetheless has a finite transition temperature, while the  D=2+1
antiferromagnetic Heisenberg model does not. One thus may raise the question whether this difference can be seen in the perturbative
effective Lagrangian approach. The answer is negative, since the physics near the finite Kosterlitz-Thouless transition temperature of the
D=2+1 quantum XY model occurring at $T_{KT} \approx 0.343 J$ (square lattice), is beyond the reach of the effective Lagrangian method. The
effective approach is only valid in the regime where the spin waves are the relevant degrees of freedom: low temperature and weak external
field. The effective series are on safe grounds up to maybe one third of the topological transition temperature $T_{KT}$, but they
definitely break down as one approaches $T_{KT}$ from below. Hence the effective method cannot see this difference between the XY model
and the antiferromagnet in two spatial dimensions.

\section{Conclusions}
\label{Summary}

Below the Kosterlitz-Thouless phase transition, which takes place at $T_{KT} \approx 0.343 J$ for the square lattice, the physics of the
d=2+1 quantum XY model is dominated by the spin waves.
Within the effective Lagrangian perspective, we have analyzed the partition function up to three loop-order and have derived the
low-temperature series for various thermodynamic quantities, including the order parameter.

Although the low-temperature regime of the d=2+1 quantum XY model has been explored before, here, for the first time, to the best of our
knowledge, we have performed a fully systematic study by using effective field theory. In particular, we have discussed how the spin-wave
interaction manifests itself at low temperatures.

In the pressure, the interaction shows up at next-to-leading order through a term proportional to four powers of the temperature, related
to a two-loop graph. Subsequent contributions originate from three-loop graphs and are of order $T^5$. We also pointed out that, in the
case of the d=2+1 Heisenberg antiferromagnet, the two-loop contribution is zero: the spin-wave interaction only sets in at order $T^5$.
Still, as we have explained, the strength of the interaction is larger in the Heisenberg antiferromagnet than in the quantum XY model. In
both cases, the interaction is repulsive. While the spin-wave interaction in the d=2+1 quantum XY model tends to zero for very weak
staggered field, in the d=2+1 Heisenberg antiferromagnet, on the other hand, the spin-wave interaction gets stronger.

It is essential that the external staggered field $H_s$ in our effective analysis is kept finite. Switching it off completely, our
effective series become meaningless, because the "critical" temperature $T_c$, below which the spin-wave picture is valid, tends to zero.
This is how the Mermin-Wagner theorem raises its head in our effective calculation.

As we have argued, lattice anisotropies only become relevant in the subleading Lagrangian ${\cal L}^4_{eff}$, giving rise to a
next-to-next-to-leading order effect in the partition function, which does not affect our conclusions. Still, the lattice geometry does
affect our results in a trivial way, because the low-energy constants $\Sigma_s$ (staggered magnetization at zero temperature and zero
external field), $v$ (spin-wave velocity) and $F^2$ (spin stiffness or helicity modulus) take different values on e.g. the square and the
honeycomb lattice.

We find it quite remarkable that all these results follow from symmetry considerations only. In particular, the fact that the spin-wave
interaction in the pressure is repulsive, is an immediate consequence of the spontaneously broken symmetry O(2) ($T=0$) of the d=2+1
quantum XY model.

\section*{Acknowledgments}
The author would like to thank C. J. Hamer, F. Niedermayer, J. Oitmaa and U.-J. Wiese for useful comments on the manuscript.

\begin{appendix}

\section{Quantum XY Model versus Heisenberg Model}
\label{appendixA}

In this Appendix we point out that the effective descriptions of the quantum XY model and the Heisenberg model are different, although
both systems are characterized by a spontaneously broken rotation symmetry
at zero temperature.
In particular, while there exists a mapping between the ferromagnetic and the antiferromagnetic quantum XY model also on the effective
level, the Heisenberg ferromagnet is rather special.

On the microscopic level, the Heisenberg model,
\begin{equation}
{\cal H}_0 = -J \sum_{n.n.} {\vec S}_m \cdot {\vec S}_n ,
\end{equation}
is invariant under internal spin O(3) rotations. The ferromagnetic ($J>0$) or antiferromagnetic ($J<0$) ground states, however, are
invariant only under O(2). The well-known differences between ferromagnetic and antiferromagnetic spin waves -- the former present a
quadratic dispersion law, the latter follow a linear relation -- have been analyzed from a unified perspective based on symmetries within
the effective Lagrangian framework in Ref.~\citep{Leu94a}. The difference between the two systems can be traced back to the question
whether or not the expectation value of the charge densities $J^0_i$ of the spin rotation symmetry is zero. The vacuum expectation value
of $J^0_i$ is given by the spontaneous magnetization $\Sigma$,
\begin{equation}
\langle 0 | J^0_i | 0 \rangle = \delta^3_i \Sigma , \quad i = 1,2,3 ,
\end{equation}
pointing here along the direction of the third axis. At leading order in the effective description of the ferromagnet, the spontaneous
magnetization shows up as the coefficient of a topological term which involves only a single time derivative and dominates the low-energy
properties of the system. More precisely, the leading-order effective Lagrangian of the Heisenberg ferromagnet takes the form (see
Ref.~\citep{Leu94a})
\begin{equation}
\label{LeffFerro}
{\cal L}^2_{eff}[F] = \Sigma \frac{\epsilon_{ab} {\partial}_0 U^a U^b}{1+ U^3}
- \mbox{$\frac{1}{2}$} F^2 {\partial}_r U^i {\partial}_r U^i , \quad i = 1,2,3 .
\end{equation}
It also contains a term with two spatial derivatives, proportional to the square of the low-energy constant $F$. Ferromagnetic spin-waves
in the Heisenberg model thus obey a quadratic dispersion law.

On the other hand, since the spontaneous magnetization is zero for the Heisenberg antiferromagnet, the topological term with just one time
derivative is absent. In the leading-order effective Lagrangian,
\begin{equation}
\label{LeffAF}
{\cal L}^2_{eff}[AF] = \mbox{$\frac{1}{2}$} F^2_1 {\partial}_0 U^i{\partial}_0 U^i
- \mbox{$\frac{1}{2}$} F^2_2 {\partial}_r U^i{\partial}_r U^i , \quad i = 1,2,3 ,
\end{equation}
time and space derivatives are one the same footing. Antiferromagnetic spin waves follow a linear, i.e., relativistic dispersion law, with
the velocity of light replaced by the spin-wave velocity $v=F_2/F_1$. Setting $v \equiv 1$, we may use relativistic notation
\begin{eqnarray}
\label{LeffAFrel}
{\cal L}^2_{eff}[AF] & = & \mbox{$ \frac{1}{2}$} F^2 \partial_{\mu} U^i \partial^{\mu} U^i , \quad F_1 = F_2 = F , \nonumber \\
& & \qquad i = 1,2,3 ,
\end{eqnarray}
where (pseudo-)Lorentz invariance is manifest. The essential point is to realize that the coefficient of the topological term, i.e. the
spontaneous magnetization, makes the difference between Heisenberg ferromagnets and antiferromagnets on the effective level.

As pointed out in Ref.~\citep{Leu94a}, the topological term can only arise when the spontaneously broken symmetry is non-Abelian.
Therefore, despite the fact that the ferromagnetic XY model is also characterized by a nonzero spontaneous magnetization, the topological
term is absent. At leading order in the effective description, there is thus no difference between the XY ferromagnet and the XY
antiferromagnet. In either case the effective Lagrangian, using relativistic notation, is given by
\begin{equation}
\label{LeffXY}
{\cal L}^2_{eff}[XY] = \mbox{$ \frac{1}{2}$} F^2 \partial_{\mu} U^i \partial^{\mu} U^i , \quad i = 1,2 .
\end{equation}
This is equivalent to the statement that, in the absence of external fields, the ferromagnetic and antiferromagnetic quantum XY models can
be mapped onto each other by a unitary transformation. Note that, apart from the number of magnon fields and the actual value of the
low-energy coupling $F$, the effective Lagrangian for the quantum XY model (\ref{LeffXY}) coincides with the effective Lagrangian
(\ref{LeffAFrel}) for the Heisenberg antiferromagnet.

We now consider the incorporation of external fields. On the microscopic level, in the Heisenberg model, one may introduce a magnetic
field $\vec H = (0,0,H)$ that couples to the magnetization vector $\sum_n {\vec S}_n$, and a staggered field ${\vec H}_s = (0,0,H_s)$ that
couples to the staggered magnetization vector $\sum_n (-1)^n {\vec S}_n$,
\begin{equation}
\label{HeisenbergExtension}
{\cal H} = {\cal H}_0 - \sum_n {\vec S}_n \cdot {\vec H} - \sum_n (-1)^n {\vec S}_n \cdot {\vec H_s} .
\end{equation}
Again we assume that the underlying lattice is bipartite. On the effective level, the leading-order Lagrangians for the ferromagnet and
the antiferromagnet then take the form \citep{Hof99a}:
\begin{eqnarray}
\label{LeffFerroAFHp2}
{\cal L}^2_{eff}[F,H] & = & \Sigma \frac{\epsilon_{ab} {\partial}_0 U^a U^b}{1+ U^3}
- \mbox{$\frac{1}{2}$} F^2 {\partial}_r U^i {\partial}_r U^i \nonumber \\
& + & \Sigma H^i U^i , \quad i = 1,2,3 , \\
{\cal L}^2_{eff}[AF,H_s,H] & = & \mbox{$\frac{1}{2}$} F^2 D_{\mu} U^i D^{\mu} U^i + {\Sigma}_s H_s^i U^i ,  \nonumber \\
& &  i = 1,2,3 .
\end{eqnarray}
As shown in Ref.~\citep{Leu94a}, the magnetic field enters the effective Lagrangian ${\cal L}^2_{eff}[AF,H_s,H]$ in the time component of
the covariant derivative of $\vec U$,
\begin{equation}
\label{CovariantH}
D_0 U^i = \partial_0 U^i + \varepsilon_{ijk} H^j U^k .
\end{equation}
Note that we do not consider the case where the staggered field is introduced for the ferromagnet.

At next-to-leading order, in the antiferromagnetic Lagrangian, the magnetic field $\vec H$ shows up again in the covariant derivative
$D_0$,
\begin{equation}
{\cal L}^4_{eff}[AF,H] = e_1 (D_{\mu} U^i D^{\mu} U^i)^2 + e_2 (D_{\mu} U^i D^{\nu} U^i)^2 ,
\end{equation}
where the quantities $e_1$ and $e_2$ are two additional low-energy constants. The ferromagnet, however, is rather special in the sense
that the time derivatives, along with the magnetic field, can be eliminated with the equation of motion \citep{Hof02}, such that the
next-to-leading order Lagrangian takes the form
\begin{eqnarray}
{\cal L}^4_{eff}[F,H] & = & l_1 {( {\partial}_r U^i {\partial}_r U^i )}^2 \nonumber \\
& + & l_2 {( {\partial}_r U^i {\partial}_s U^i )}^2 + l_3 \Delta U^i \Delta U^i .
\end{eqnarray}
Finally, the staggered field $\vec H_s$ gives rise to the following three extra terms in the next-to-leading-order Lagrangian of the
antiferromagnet \citep{HL90}:
\begin{eqnarray}
\label{LeffAFStaggered}
{\cal L}^4_{eff}[AF,H_s] & = & k_1 \frac{{\Sigma}_s}{F^2} (H_s^i U^i) ({\partial}_{\mu} U^k {\partial}^{\mu} U^k) \nonumber \\
& + & k_2 \frac{{\Sigma}_s^2}{F^4} (H_s^i U^i)^2 + k_3 \frac{{\Sigma}_s^2}{F^4} H_s^i H_s^i .
\end{eqnarray}
Gathering partial results, the effective Lagrangians up to order $p^4$ for the Heisenberg ferromagnet and antiferromagnet, in the presence
of external fields, amount to
\begin{eqnarray}
{\cal L}^F_{eff}[H] & = & \Sigma \frac{\epsilon_{ab} {\partial}_0 U^a U^b}{1+ U^3}
- \mbox{$\frac{1}{2}$} F^2 {\partial}_r U^i {\partial}_r U^i + \Sigma H^i U^i \nonumber \\
& + & l_1 {( {\partial}_r U^i {\partial}_r U^i )}^2 + l_2 {( {\partial}_r U^i {\partial}_s U^i )}^2 \nonumber \\
& + & l_3 \Delta U^i \Delta U^i , \quad i = 1,2,3 ,
\end{eqnarray}
and
\begin{eqnarray}
{\cal L}^{AF}_{eff}[H,H_s] & = & \mbox{$ \frac{1}{2}$} F^2 D_{\mu} U^i D^{\mu} U^i + {\Sigma}_s H_s^i U^i \nonumber \\
& + & e_1 (D_{\mu} U^i D^{\mu} U^i)^2 + e_2 (D_{\mu} U^i D^{\nu} U^i)^2 \nonumber \\
& + & k_1 \frac{{\Sigma}_s}{F^2} (H_s^i U^i) (D_{\mu} U^k D^{\mu} U^k) + k_2 \frac{{\Sigma}_s^2}{F^4} (H_s^i U^i)^2 \nonumber \\
& + & k_3 \frac{{\Sigma}_s^2}{F^4} H_s^i H_s^i , \quad i = 1,2,3.
\end{eqnarray}

We now turn to the quantum XY model. Interestingly, in the case of the Abelian symmetry O(2), the term involving the magnetic field
in the covariant derivative $D_0$ (\ref{CovariantH}) vanishes. Hence the only field that matters for the XY antiferromagnet is the
staggered field ${\vec H}_s = (0,H_s)$ and we have
\begin{eqnarray}
{\cal L}^{XY,AF}_{eff}[H_s] & = & \mbox{$ \frac{1}{2}$} F^2 {\partial}_{\mu} U^i {\partial}^{\mu} U^i + {\Sigma}_s H_s^i U^i \nonumber \\
& + & e_1 (\partial_{\mu} U^i {\partial}^{\mu} U^i)^2 + e_2 (\partial_{\mu} U^i {\partial}^{\nu} U^i)^2 \nonumber \\
& + & k_1 \frac{{\Sigma}_s}{F^2} (H_s^i U^i) ({\partial}_{\mu} U^k {\partial}^{\mu} U^k) \nonumber \\
& + & k_2 \frac{{\Sigma}_s^2}{F^4} (H_s^i U^i)^2 + k_3 \frac{{\Sigma}_s^2}{F^4} H_s^i H_s^i , \nonumber \\
& & i = 1,2.
\end{eqnarray}
In the case of the ferromagnet, again, the topological term involving just one time derivative can only occur if the spontaneously broken
symmetry is non-Abelian. As a consequence, in the effective Lagrangian of the XY ferromagnet, unlike for the Heisenberg ferromagnet, time
and space derivatives are on the same footing. The equation of motion no longer is of first order in the time derivative, such that time
derivatives can no longer be eliminated. Finally, the magnetic field couples to the spontaneous magnetization vector $\vec U$, i.e., it
enters the effective Lagrangian in the same way as the staggered field which couples to the staggered magnetization vector $\vec U$ in
the case of the XY antiferromagnet, namely
\begin{eqnarray}
\label{XYferro}
{\cal L}^{XY,F}_{eff}[H] & = & \mbox{$ \frac{1}{2}$} F^2 {\partial}_{\mu} U^i {\partial}^{\mu} U^i + {\Sigma} H^i U^i \nonumber \\
& + & e_1 (\partial_{\mu} U^i {\partial}^{\mu} U^i)^2 + e_2 (\partial_{\mu} U^i {\partial}^{\nu} U^i)^2\nonumber \\
& + & k_1 \frac{\Sigma}{F^2} (H^i U^i) ({\partial}_{\mu} U^k {\partial}^{\mu} U^k) + k_2 \frac{{\Sigma}^2}{F^4} (H^i U^i)^2 \nonumber \\
& + & k_3 \frac{{\Sigma}^2}{F^4} H^i H^i , \quad i = 1,2.
\end{eqnarray}
In conclusion, the systematic effective field theory analysis for the quantum XY model in the presence of magnetic and staggered fields
just reflects what is known from the microscopic analysis. On a bipartite lattice, there is a one-to-one correspondence between the XY
ferromagnet in a magnetic field, and the XY antiferromagnet in a staggered field.

\section{Cateye Graph in D=2+1: Evaluation}
\label{appendixB}

In the case of the antiferromagnetic Heisenberg model ($N$=3), the renormalization and subsequent numerical evaluation of the relevant
integral $J_2$ ,
\begin{equation}
J_2 = {\int}_{{\cal T}} {\mbox{d}}^d x \Big\{ \partial_{\mu} G(x) \partial_{\mu} G(x) \Big\}^2 ,
\end{equation}
was discussed in detail in Appendix A of Ref.\citep{Hof10}. The same techniques can be applied to the integral $J_1$,
\begin{equation}
J_1 = {\int}_{{\cal T}} {\mbox{d}}^d x \Big\{ G(x) \Big\}^4 ,
\end{equation}
which matters for the quantum XY model ($N$=2). This is the subject of the present Appendix. In either case, we are dealing with the
evaluation of the cateye graph 5c of Fig.~\ref{figure1}.

We first decompose the thermal propagator, defined in Eq.~(\ref{ThermalPropagator}), into two pieces,
\begin{equation}
\label{decompositionX}
G(x) = \Delta (x) + {\bar G}(x) ,
\end{equation}
where $\Delta (x)$ is the zero-temperature propagator.

The singularities contained in $J_1$ can be removed by subtracting the following counterterms,
\begin{equation}
\label{FunctionsJbarJ3}
{\bar J}_1 = J_1 - c_1 - c_2 g_1(M,T) ,
\end{equation}
where $g_1(M,T)$ is the Bose function defined in Eq.~(\ref{BoseFunctions}). Using the method developed in Ref.~\citep{GL89}, we now
establish this result. This technique also allows us to derive expressions suitable for numerical evaluation.

The first step consists in cutting out a sphere $\cal S$ around the origin of radius $|{\cal S}| \leq \beta/2$ and split the integral
$J_1$ into two pieces,
\begin{equation}
\label{decoSphere}
J_1 = {\int}_{{\cal S}} {\mbox{d}}^d x  \Big\{ G(x) \Big\}^4
+ {\int}_{{\cal T} \setminus {\cal S}} {\mbox{d}}^d x \Big\{ G(x) \Big\}^4 .
\end{equation}
The integrand in the second term, involving an integration over ${\cal T} \setminus {\cal S}$, is not singular, such that the limit
$d\to3$ poses no problems. In the first expression, where the integration extends over the sphere, we use the decomposition
(\ref{decompositionX}) and arrive at
\begin{equation}
\label{insertion}
{\int}_{\cal S} {\mbox{d}}^d x \Big( { {\bar G} }^4
+ 4 { {\bar G} }^3 \Delta
+ 6 { {\bar G} }^2 {\Delta}^2
+ 4 {\bar G} {\Delta}^3
+ {\Delta }^4 \Big) .
\end{equation}
While the first three terms are convergent in $d$=2+1, the other two are divergent. The singularities contained therein can be taken care
of as follows \citep{GL89}. The quantity $\Delta (x)$ is Euclidean invariant. Therefore the expression
\begin{equation}
{\int}_{{\cal S}} {\mbox{d}}^d x 4 {\bar G} \Delta^3
\end{equation}
merely involves the angular average of ${\bar G} (x)$,
\begin{equation}
f(R) = {\int} {\mbox{d}}^{d-1} \Omega {\bar G}(x) , \quad R = |x| .
\end{equation}
The function ${\bar G} (x)$ obeys the differential equation
\begin{equation}
\Box {\bar G} = M^2 {\bar G} ,
\end{equation}
which implies
\begin{equation}
\Bigg( \frac{\mbox{d}^2}{\mbox{d} R^2} + \frac{d-1}{R} \frac{\mbox{d}}{\mbox{d} R} - M^2 \Bigg) f = 0 , \quad R < \beta .
\end{equation}
Now the function $g_1 ch(M x_4)$ satisfies the same differential equation as ${\bar G}(x)$ and agrees with it at the origin. One thus
concludes that the angular averages of the two quantities are identical:
\begin{equation}
{\int}_{{\cal S}} {\mbox{d}}^d x {\bar G} {\Delta}^3
= g_1 {\int}_{{\cal S}} {\mbox{d}}^d x ch(M x_4) {\Delta}^3 .
\end{equation}

Finally, decomposing the integral over the sphere according to
\begin{eqnarray}
& & 4 g_1 {\int}_{{\cal S}} {\mbox{d}}^d x ch(M x_4) {\Delta}^3 =  4 g_1 {\int}_{{\cal R}} {\mbox{d}}^d x ch(M x_4) {\Delta}^3 \nonumber \\
& & - 4 g_1 {\int}_{{\cal R} \setminus {\cal S}} {\mbox{d}}^d x  ch(M x_4) {\Delta}^3 ,
\end{eqnarray}
the singularity now shows up in the integral over all Euclidean space,
\begin{equation}
c_2 = 4 {\int}_{{\cal R}} {\mbox{d}}^d x ch(M x_4) {\Delta}^3 .
\end{equation}

Turning to the last expression in Eq.~(\ref{insertion}), we remove the singularity by subtracting the temperature-independent integral of
${\Delta(x)}^4$ over $\cal R$,
\begin{equation}
c_1 = {\int}_{{\cal R}} {\mbox{d}}^d x { \Delta }^4 .
\end{equation}
Gathering partial results, the renormalized integral ${\bar J}_1$ in $d$=2+1 can be written as
\begin{eqnarray}
\label{J1bar}
{\bar J}_1 & = & {\int}_{{\cal T}} {\mbox{d}}^3 x T
+ {\int}_{{\cal T} \setminus {\cal S}} {\mbox{d}}^3 x U
- {\int}_{{\cal R} \setminus {\cal S}} {\mbox{d}}^3 x W , \nonumber \\
T & = & { {\bar G} }^4 + 4 { {\bar G} }^3 \Delta + 6 { {\bar G} }^2 { \Delta }^2 , \nonumber \\
U & = & 4 {\bar G} { \Delta }^3 + { \Delta }^4 , \nonumber \\
W & = & 4 g_1 ch(M x_4) { \Delta }^3 + { \Delta }^4 .
\end{eqnarray}
All pieces in the above representation of ${\bar J}_1$ are now finite. Since the quantities ${\bar G}(x)$ and $\Delta(x)$ only depend on
$r=|{\vec x}|$ and on $t=x_4$, the above integrals are in fact two-dimensional
\begin{equation}
{\mbox{d}}^3 x = 2 \pi r {\mbox{d}} r {\mbox{d}} t .
\end{equation}
Note that the size of the sphere, $|{\cal S}| \leq \beta/2$, introduced in the decomposition (\ref{decoSphere}), is arbitrary and that the
quantity ${\bar J}_1$ cannot depend on the specific radius of the sphere we choose. This fact allows us to check our numerical results.
Indeed, we have verified that, using different sizes of the sphere, we arrive at the same result for the function $q(\sigma)$, which can
be extracted from ${\bar J}_1$ through its definition (\ref{definitionI}).

\end{appendix}

\end{document}